\documentclass[11pt,a4paper]{article}
\usepackage{authblk}

\usepackage{graphicx, xcolor, hyperref}
\usepackage{amsmath}
\usepackage{epsf}
\usepackage{bm, ulem}
\usepackage{braket}
\usepackage{cancel}
\usepackage{times}
\usepackage{mathrsfs}
\usepackage{amssymb}
\usepackage{epsfig}
\usepackage[margin=1.25in]{geometry}
\usepackage{tikz}
\usepackage{smartdiagram}
\usetikzlibrary{arrows}
\usetikzlibrary{decorations.pathreplacing}
\usepackage{comment}

\input epsf.sty

\def\doi{http://dx.doi.org/}

\newcommand{\be}{\begin{equation}}
\newcommand{\ee}{\end{equation}}
\newcommand{\bea}{\begin{eqnarray}}
\newcommand{\eea}{\end{eqnarray}}

\begin{document}

\author[1]{Ivar Lyberg\thanks{ivar\_lyberg@hotmail.com}}
\author[2]{Vladimir Korepin}
\author[3]{Jacopo Viti\thanks{jacopo.viti@fi.infn.it (corresponding author)}}
\affil[1]{ PhD Stony Brook University}
\affil[2]{C.N. Yang Institute for Theoretical Physics, Stony Brook University, Stony Brook, New York, 11794-3840, USA}
\affil[3]{INFN, sezione di Firenze, via G. Sansone 1,
50019 Sesto Fiorentino, Italy}
\title{Fluctuation of the phase boundary in the six-vertex model with Domain Wall Boundary Conditions: a Monte Carlo study}
\date{}

\maketitle

\begin{abstract}
We consider the six-vertex model with Domain Wall Boundary Conditions in a  square lattice of dimension $N\times N$. Our main interest is the study of the fluctuations of the extremal lattice path about the arctic curves.  We address the problem through Monte Carlo simulations.  At  $\Delta = 0$, the fluctuations of the extremal path along any line parallel to the square diagonal were rigorously proven to follow the Tracy-Widom distribution. We provide strong numerical evidence that this is true also for other values of the anisotropy parameter $\Delta$ ($0\leq \Delta < 1$). We argue that the typical width of the fluctuations of the extremal path about the arctic curves scales as  $N^{1/3}$ and provide a numerical estimate for the parameters of the scaling random variable.
\end{abstract}

\section{Introduction}
\label{sec:intro}

In probability theory~\cite{Feller},  the sum of a large number of random variables once divided by the number of summands might approach a deterministic value. For instance, the law of large numbers implies that the sample mean of $N$ independent identically distributed random variables converges to their distribution average for large $N$. Moreover, if those random variables have finite variance, fluctuations of their sum are normally distributed with a variance of order $\sqrt{N}$. The convergence toward smooth functions of combinations of a large number of random variables is at the root of the effectiveness of the methods of statistical mechanics and also provides a solid ground for numerical techniques such as Monte Carlo. This phenomenon is not restricted to local observables but extends to geometrical quantities such as random curves. An instructive example discussed in~\cite{Ok} (but see also~\cite{CS2016}) is a one-dimensional random walk starting from the origin and conditioned in $T$ steps to reach the point $X$. For $X,~T\rightarrow\infty$ with $X/T=v$ finite, the random paths performed by the walker are with probability one within any neighborhood of the classical trajectory $X=vT$. Limit shapes~\cite{KO} are smooth nonrandom curves that describe the thermodynamic limit of random lattice paths. Historically, one of the most studied examples is the arctic circle~\cite{PROPP1998} that appears in dimer coverings of the Aztec diamond. This domino tiling problem is equivalent~\cite{FS} to the six vertex model~\cite{P35, LIEB} with Domain Wall Boundary Conditions (DWBC)~at its free fermionic point~\cite{BAXTER, KBI,  RES_lect}.

\begin{figure}
\centering
\includegraphics{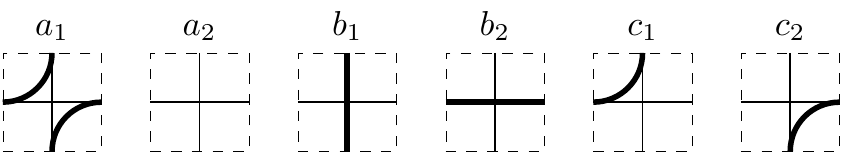}
\caption{The six possible configurations of a lattice path in the six-vertex model are denoted by thick lines. The continuous lines are lattice edges while dashed lines are edges of the dual lattice.}
\label{fig:conf}
\end{figure}

Let $N$ be a positive integer, a configuration of the six vertex model on a $N\times N$ square lattice with DWBC~\cite{Korepin1982, Izergin, ICK, zj2000, eloranta1999, kzj} is specified by $N$ non-intersecting paths traveling on the edges of the lattice. The lines enter from above and exit on the left. On each vertex, the path can have the six possible configurations given in Fig.~\ref{fig:conf}; the Boltzmann weights associated with those are $a$ for the vertices of type $a_1$ and $a_2$, $b$~for the vertices of type $b_1$ and $b_2$ and $c$~for the vertices of type $c_1$ and $c_2$. The phase diagram of the model is characterized by  the ratio $b/a$ and the parameter $\Delta$~\cite{BAXTER}, defined as
$$
\Delta:=\frac{a^2+b^2-c^2}{2ab}.
$$  
In this work, we will focus on the domain $0\leq\Delta<1$, which is included in the so-called disordered phase $|\Delta|<1$ see~\cite{BLbook} for a comprehensive study of the phase diagram. The case $\Delta=0$ is the free fermionic point mentioned a few lines above.

\begin{figure}
\centering
\includegraphics[width=0.5\textwidth]{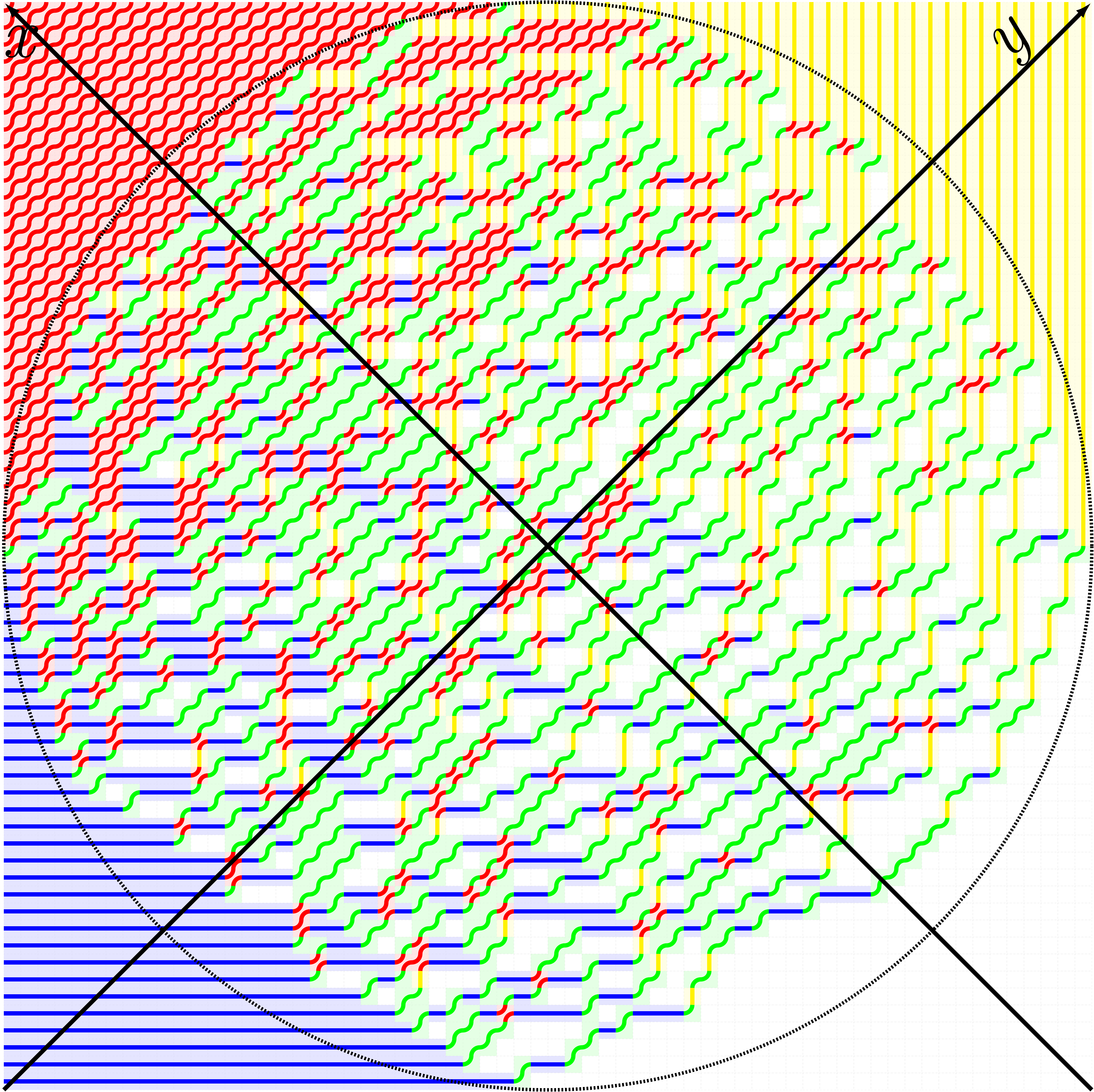}~~~
\includegraphics[width=0.5\textwidth]{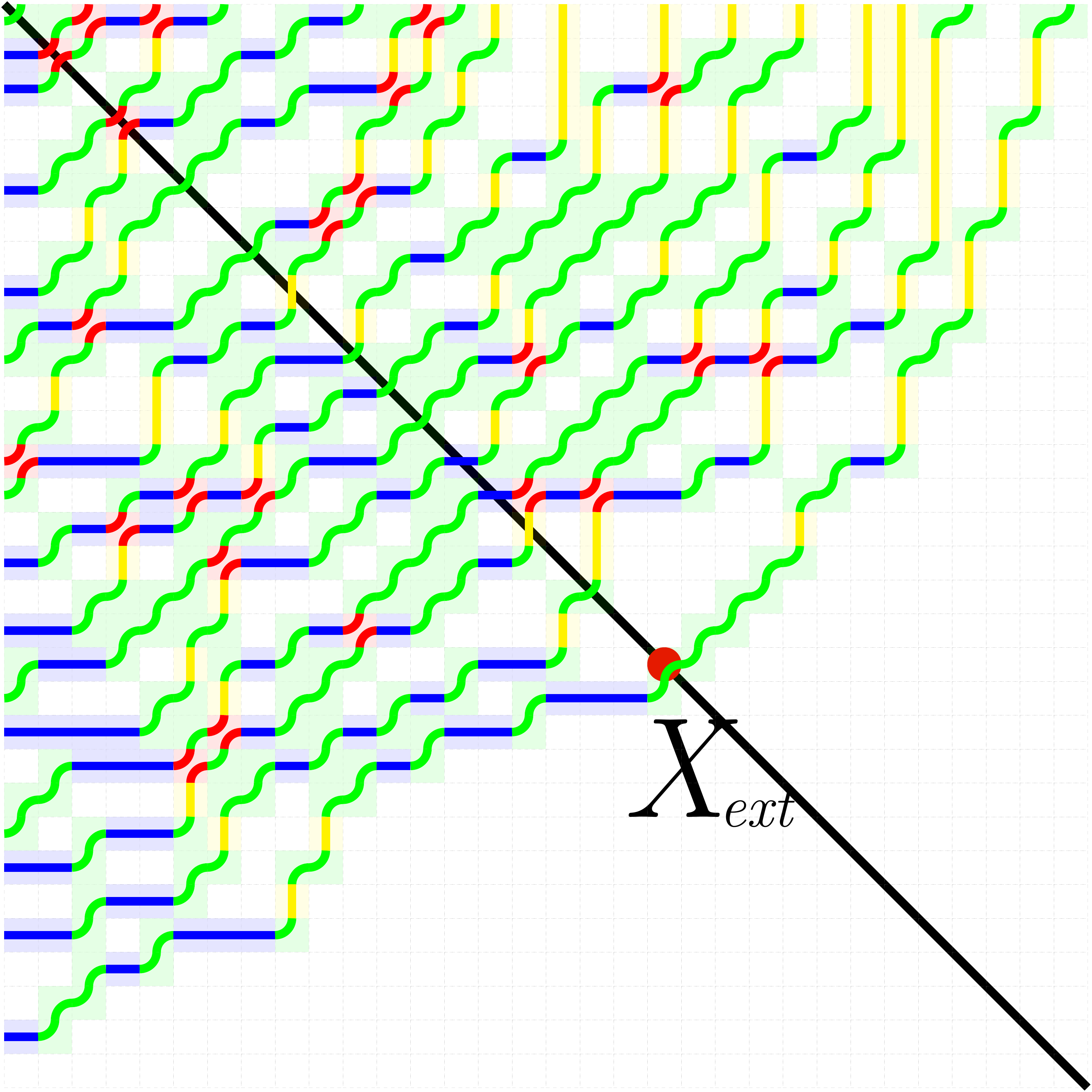}
\caption{\textbf{Left.} Lattice path configuration in the six vertex model with DWBC for $N=64$, and $a=b$ at $\Delta=0$. The possible vertex configurations have been colored differently: vertices of type $a_1$ are colored in red, of type $a_2$ in white, of type $b_1$ in yellow, of type $b_2$ in blue while of type $c_1$ and $c_2$ in green. The black circle is the arctic curve for the given choice of the parameters. \textbf{Right.} The SE corner of the lattice path configuration on the left for $N=64$. The furthest path from the center of the lattice intersects a constant $y$ line in a point with abscissa $X_{ext}(\eta)$. The picture shows $X_{ext}(\eta=0)$. }
\label{fig:extermal}
\end{figure}

\noindent
 The $N$ random paths cannot intersect and for  large enough $N$ they can fluctuate only within a region whose boundary, after rescaling all the distances by $N$, is arbitrarily close to a curve, dubbed the arctic curve~\cite{PROPP1998}. For $\Delta=0$ and $a=b$, such a curve is the circle $(x/R)^2+(y/R)^2=1$, depicted in Fig.~\ref{fig:extermal} and inscribed in the square lattice. Inside the arctic curve the model is in a disordered phase, outside is in a trivial ferromagnetic phase. The Cartesian coordinates are also defined in Fig.~\ref{fig:extermal} and lengths, such as the circle radius $R$, are measured in suitable lattice units.  In this paper, we follow the convention to divide the square diagonal into $N$ intervals, therefore $x,y$ are semi-integers in the interval $[-N/2, N/2]$; for simplicity, we assume $N$ a power of two from now on.
 
 The Cartesian equation of the arctic curve depends on the parameter $\Delta$ and was calculated in~\cite{PC, PCZ}, see also~\cite{CS2016}. A rigorous derivation at $\Delta=1/2$ has been given in~\cite{Agra}.
 
 The limit shape, however, is a property of the model in the thermodynamic limit and it is relevant to understand how the interface between the disorder and the ferromagnetic phase fluctuates due to finite-size effects. In this case, the phases are not separated by a smooth curve; rather, the distance from a corner to the disordered phase varies according to a probability distribution. Consider, for instance, a quarter of the original lattice: the southeast (SE) quadrant on the right of Fig.~\ref{fig:extermal}. When $N$ is large, there are no paths in a neighborhood of the SE corner and all the vertices are of type $a_2$. For later purposes, introduce the parameter $\eta$ as 
 \begin{equation}
 \eta:=2y/N.
 \label{eq:etadef}
 \end{equation}
 Take $|\eta|<1/2$; as we move along a constant-$y$  line toward the boundary (for $x>0$), we cross $N/4$  lattice paths, see Fig.~\ref{fig:extermal}. Denote then by $X_{ext}(\eta)$ the $x$-coordinate of the intersection of a constant-$y$ line with the furthest path from the origin. Refs.~\cite{kj2, kj} tackled rigorously the study of the fluctuations of the extremal path in the context of dimer coverings. The same conclusions apply to the six-vertex~ with DWBC at $\Delta=0,~$\cite{FS}. In particular, it can be proven that for $N\gg 1$,
  \begin{equation}
  \label{eq:chi_def}
  X_{ext}(\eta)\stackrel{N\gg 1}{\longrightarrow}\Lambda(\eta) N \pm \Gamma(\eta) N^{1/3}\mathcal{Z},
  \end{equation}
where $\mathcal{Z}$ follows a Tracy-Widom (TW) distribution~\cite{TW} for any $|\eta|<1/2$. The sign in front of Eq.~\eqref{eq:chi_def}, is the one of $X_{ext}(\eta)$ in the coordinate system defined in Fig.~\ref{fig:extermal}. Moreover, the parameters $\Lambda$ and $\Gamma$ are $O(1)$: $\Lambda N$ is the $x$-coordinate of the  intersection of the arctic curve (an ellipse if $a\not=b$) with the constant-$y$ line while $\Gamma$  can be derived from Ref.~\cite{ADSV2016}, see Sec.~\ref{sec:Delta0}. 
 
The TW distribution characterizes fluctuations of extrema of random variables in several different contexts, among them: the spectrum of Hermitian random matrices~\cite{TW}, ordered sequences in random permutations~\cite{BDJ}, random growth models~\cite{kj2, PS} and their generalizations~\cite{TW2, CGHS}, the KPZ~\cite{KPZ} equation~\cite{SS}, quantum dynamics~\cite{Racz, ADSV2016, RT}. Many of these problems can be mapped into each other~\cite{MS} and one might wonder~\cite{Deift} how universal is the TW distribution. Ref.~\cite{ACJ} proved for instance that the maximum reached by the extremal path at $\Delta=1/2$ follows a TW distribution (although with $\beta=1$).
 
 In this paper, we will study numerically the probability distribution of $X_{ext}(\eta)$ for values of $\Delta\not=0$ in the interval $[0,1)$. We will argue that after a rescaling analogous to Eq.~\eqref{eq:chi_def}, this random variable follows,  for large enough $N$,  again a TW distribution. This result is not unexpected and can be justified heuristically. Non-intersecting lattice paths can be mapped to fermionic trajectories in a Euclidean space time~\cite{ADSV2016}. The fermionic particles are interacting if $\Delta\not=0$, however the particle furthest from the origin, which is also the fastest, behaves as if were free~\cite{MS, Stephan19 , Stephan}. This observation is also at the root of the so-called tangent method~\cite{CS2016}.
 
The rest of the paper is organized as follows. In Sec~\ref{sec:num}, we review briefly the numerical method. In Section~\ref{sec:Delta0} we discuss the comparison with the case $\Delta=0$ and further analyze the lattice path fluctuations for $\Delta\not=0$ in Sec.~\ref{sec:res_num}. The analysis in Sec.~\ref{sec:res_num} focuses on the lattice diagonal ($\eta=0$), while some comments on results at $\eta\not=0$ are contained in the two conclusive appendices. Finally, in Sec.~\ref{conclusion} we summarize our conclusions.

  \section{Monte Carlo study of the six-vertex model with DWBC}
\label{sec:num}
\begin{figure}[t]
\includegraphics[width=\textwidth]{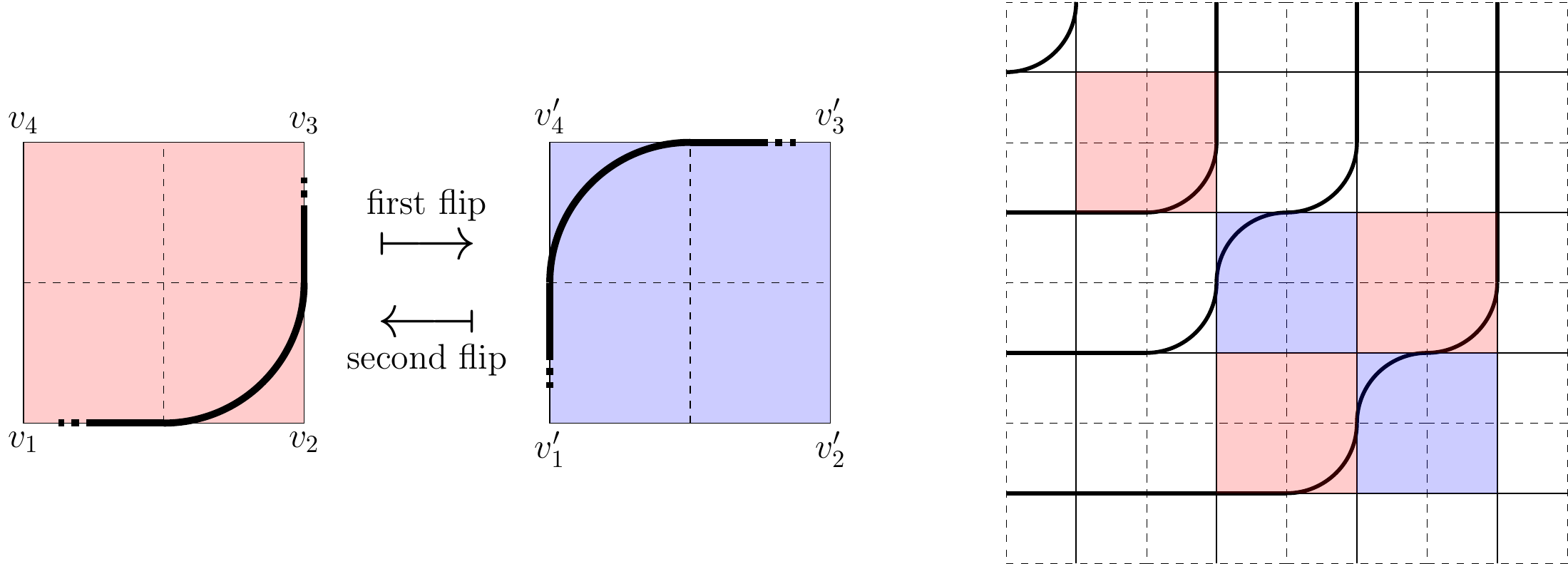}
\caption{\textbf{Left.} The two MC moves.  \textbf{Right.} The six-vertex model with DWBC with $N=4$. The picture also shows an allowed state. This state has four flippable plaquettes. The plaquettes  indicated in blue, are second flippable while the plaquettes indicated  in red are first flippable.}
\label{dwbcfig}
\end{figure}

\textit{Method.---}{We modify  a Monte Carlo (MC) algorithm for simulating the six-vertex model with DWBC  that was originally proposed in~\cite{AR2005}.} 
The MC algorithm involves two moves
called first flip and second flip; see Fig. \ref{dwbcfig}. The moves modify the configuration of the paths along the edges of a plaquette of the square lattice. The state $S$ of a plaquette is a list that contains the configurations of its four vertices: $S=\{v_1,v_2,v_3,v_4\}$, with  $v_i$ given in Fig.~\ref{fig:conf}. The Boltzmann weight of $v_i$ will be denoted by $w(v_i)=a,b,~c$. The weight of the state $S$ is then $W(S)=\prod_{i=1}^4w(v_i)$.  
An MC move takes then a state $S$ to a new state $S'$. Since in a flippable plaquette, each vertex can have two possible configurations, there are sixteen distinct plaquette states which are respectively first or second flippable. They have
up to nine distinct weights.

The computer code has two lists, each  containing the plaquettes flippable in one of the two different ways.  We can formulate the algorithm as follows.
First, randomly search among all plaquettes on the lattice, until a plaquette contained in one of the two lists is found.
A flip will then be attempted, and if the attempt is successful, then both lists will be updated: at the plaquette itself and at the four plaquettes sharing a side with it. The probability $P$ of an attempted flip that modifies the state of the plaquette $S$ into $S'$  is the Boltzmann weight of the new state, namely
\be P = W(S').
\label{eq1}
\ee
In the computer program, the largest of the three weights $a$, $b$, and $c$ is 1 and therefore Eq.~(\ref{eq1}) is well defined.
 The above conditions satisfy detailed balance and ensure ergodicity. An equivalent MC algorithm with Glauber dynamics has been employed in~\cite{LKV2017} and~\cite{LKRV2018} to obtain numerical estimates for the arctic curves and one-point functions of the model with several values of $\Delta$ in the disorder phase. To ensure the system has thermalized we rely then on the results of those papers. 
 We define a MC sweep as $N^2$ accepted flips.  The largest value of $N$ used in this paper is 256 whereas the longest side of the lattice used in the mentioned papers was as long as 500 units. We, therefore, consider it known at this time how long this system needs to equilibrate. From a rigorous perspective, Ref.~\cite{RanT} shows that the Glauber dynamics is rapidly mixing at $\Delta=1/2$, while Refs.~\cite{Liu, FD} proved that the Markov chain exponentially slows down for large and negative $\Delta$~\footnote{The arguments of~\cite{FD} break down in the whole disorder phase and although not rigorously proven, it is expected that Glauber dynamics is rapidly mixing for all $|\Delta|<1$.}
 
 A reliable test of  the TW distribution  is challenging in a statistical model, see for instance~\cite{MendlS} and requires a much larger number of sweeps than that employed previously in Refs.~\cite{LKV2017, LKRV2018}. In this paper, we will construct samples with  $10^6$ MC sweeps.   
 
 In the disorder phase, analogous MC algorithms are nowadays commonly adopted in numerical experiments with the six vertex model~\cite{Eloranta, KL2017,ks, D19, Ruelle1, Belov}. For earlier numerical approaches to ice models, see~\cite{barkema, syljuaasen}. Finally,  an efficient algorithm to simulate interacting dimers has been proposed in~\cite{Alet} but so far it has not been implemented in the six-vertex model.

\begin{figure}[t]
\begin{tikzpicture}
\node at (-1.25,3.25){\includegraphics[width=0.33\textwidth]{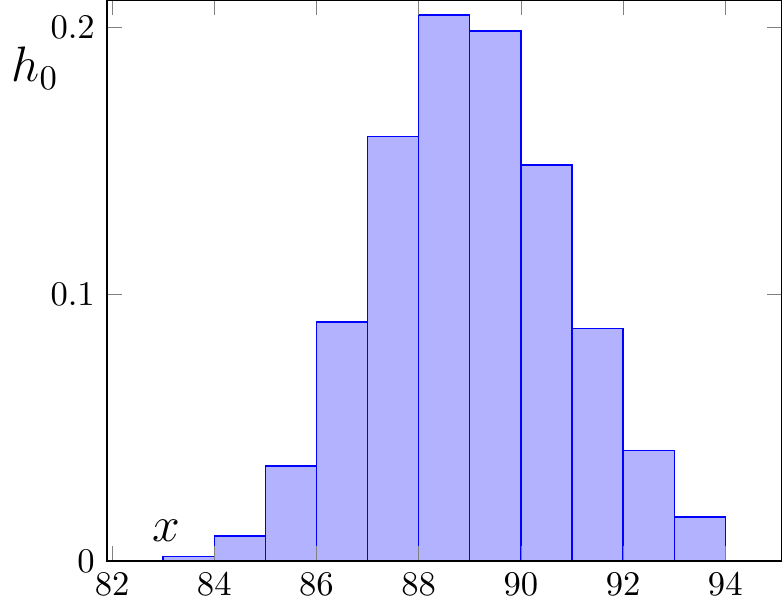}};
\node at (-1.25,-.65) {\includegraphics[width=0.35\textwidth]{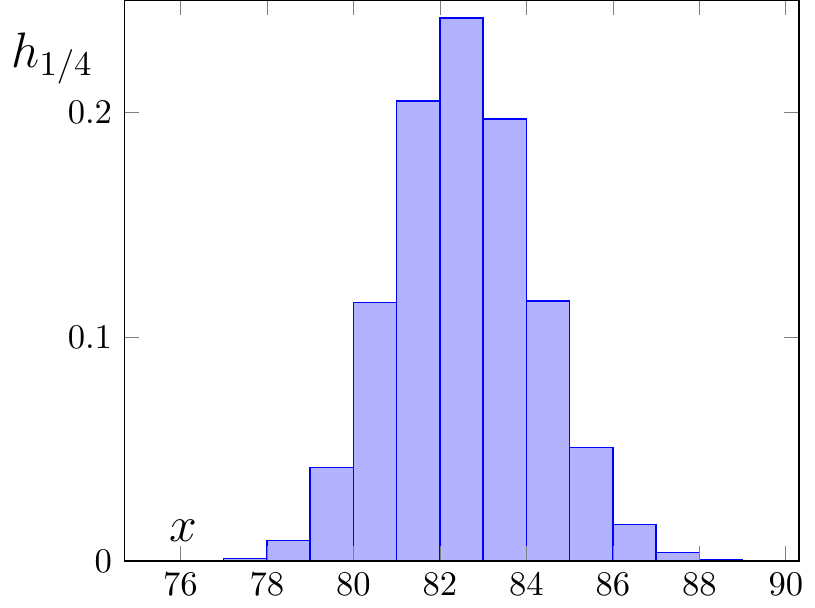}};
\node at (5.5,1.5) {\includegraphics[width=0.525\textwidth]{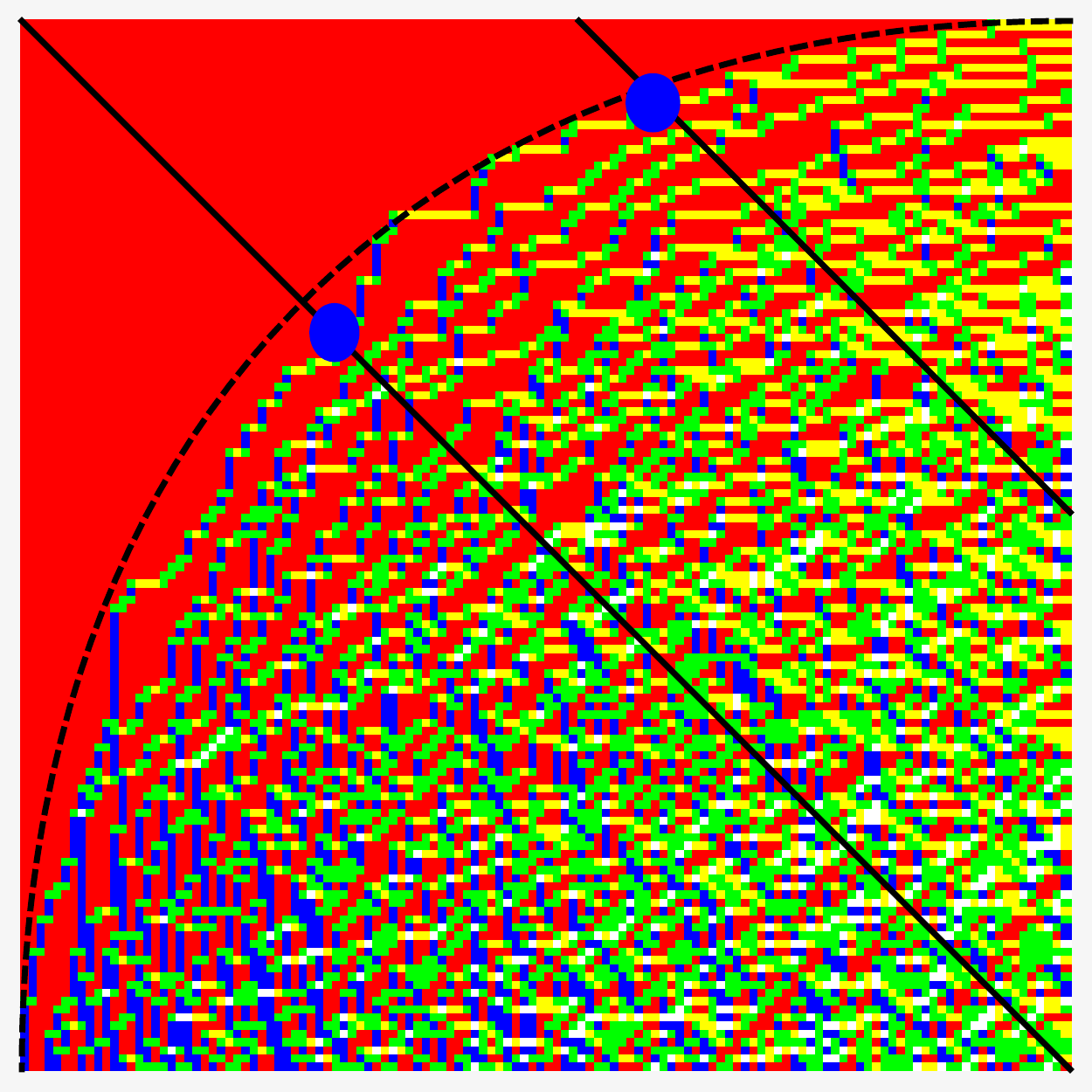}};
\node at (2.75,3) {\Large{$X_{ext}(0)$}};
\node at (4.75,4.5) {\Large{$X_{ext}(\eta)$}};
\end{tikzpicture}
\caption{\textbf{Left.} We show the numerical data for distributions of the random variables $X_{ext}(\eta=0)$ and $X_{ext}(\eta=1/4)$ along the lines $y=0$  and $y=N/8$, see also Eq.~\eqref{eq:etadef}, at $\Delta=1/2$ and $N=256$. \textbf{Right.} A realization of the random variable $X_{ext}(\eta)$ for two different values of $\eta$  at $\Delta=1/2$ and $N=256$.}
\label{fig:hist}
\end{figure}

\noindent
 \textit{The extremal lattice path.---}In the coordinate system~\footnote{Notice that in Ref.~\cite{ADSV2016}, the square diagonal is divided into $2N$ intervals. Therefore distances measured along the diagonal in this paper are twice as big as the ones considered in~\cite{ADSV2016}. This choice also implies that the constant $\Gamma$ in Eq.~\eqref{eq:chi_def} is half of the one obtained analytically in~\cite{ADSV2016} at $\Delta=0$, see Sec.~\ref{sec:Delta0}.}  defined in Fig.~\ref{fig:conf} with origin in the center of the lattice, the random variable $X_{ext}(\eta)$ is the position, along a line at constant $y$, of the last occurrence  of a vertex of different type from those in the frozen corner. The right panel of Fig.~\ref{fig:hist} illustrates a realization of  $X_{ext}(\eta)$ for $\eta=0$ and $\eta=1/4$ on the NW corner of  lattice with $N=256$ and $\Delta=1/2$. Nearby this corner the vertices are in the state $a_1$, see Fig.~\ref{fig:conf} and Fig.~\ref{fig:extermal}; the black dashed line is the arctic curve~\cite{PC}.
 
 The probability distribution, $\text{Prob}(X_{ext}(\eta)=x)$, will be approximated numerically at finite $N$  by a normalized histogram which we denote by $h_{\eta}$; see Appendix.~\ref{app_dist} for more detail about its construction.
The left panel of Fig. \ref{fig:hist} shows two examples at $\eta=0$ and $\eta=1/4$ for $\Delta=1/2$ and $N=256$. 
The two histograms are not symmetric (the right tail is longer than the left) and obviously have different variance and mean. Their properties will be studied in  the next two sections.

\section{Extremal lattice path fluctuations at  $\Delta=0$ for $|\eta|<\frac{1}{2}$}
\label{sec:Delta0}
Special attention must be paid to the case $\Delta = 0$, since it is the only point at which there are known exact results for the random process that describes the fluctuations of the coordinate $X_{ext}(\eta)$.
 Let us take for simplicity $a=b$; in~\cite{kj, kj2}, the random fluctuations were proven for large $N$ to converge to the so-called Airy process for any $|\eta|<1/2$. By following~\cite{ADSV2016}, we can derive analytically the parameters $\Gamma$ and  $\Lambda$ in Eq.~\eqref{eq:chi_def}. It turns out that
\begin{align}
\label{eq:lambda}
  & \Lambda(\eta)=\pm\frac{1}{2}\sqrt{\frac{1}{2}-\eta^2}\\
\label{eq:gamma}
  &[\Gamma(\eta)]^3=\frac{(1-4\eta^2)^2}{16(2-4\eta^2)^{3/2}},
\end{align}
where the two branches of Eq.~\eqref{eq:lambda} depend on the sign of $X_{ext}(\eta)$. Notice that $\Gamma(\eta)$ is positive and monotonically decreasing with $\Gamma(1/2)=0$. When the right-hand-side of Eq.~\eqref{eq:gamma} vanishes, the intersection point $X_{ext}(1/2)=N/4$ lies on the boundary of the square-lattice and its fluctuations are no longer described by the TW distribution. In this case, Ref.\cite{Gorin}, see also~\cite{KP} for a related discussion,  proved that the latter are Gaussian with an horizontal width $O(\sqrt{N})$.

We  then exploit Eqs.~\eqref{eq:lambda} and~\eqref{eq:gamma} to benchmark the validity of the MC simulation. In particular, according to Eq.~\eqref{eq:chi_def},  the probability distribution ($\text{Prob}(\mathcal{Z}(\eta)=\chi)$) of the random variable $\mathcal{Z}$ can be evaluated numerically as~\footnote{The notation $\lfloor x\rfloor$ indicates the integer part of $x$.}   

%%%%%%%%%%%%%%%%%%%%%%%%%%%%%%%%%%%%%%%%%%%%%%%%%%%%%%%%%%%%%%%%%%%%%%%%%%%%%%%%%%%%%%%%%%%%
\begin{figure}[t]
\includegraphics[width=.475\textwidth]{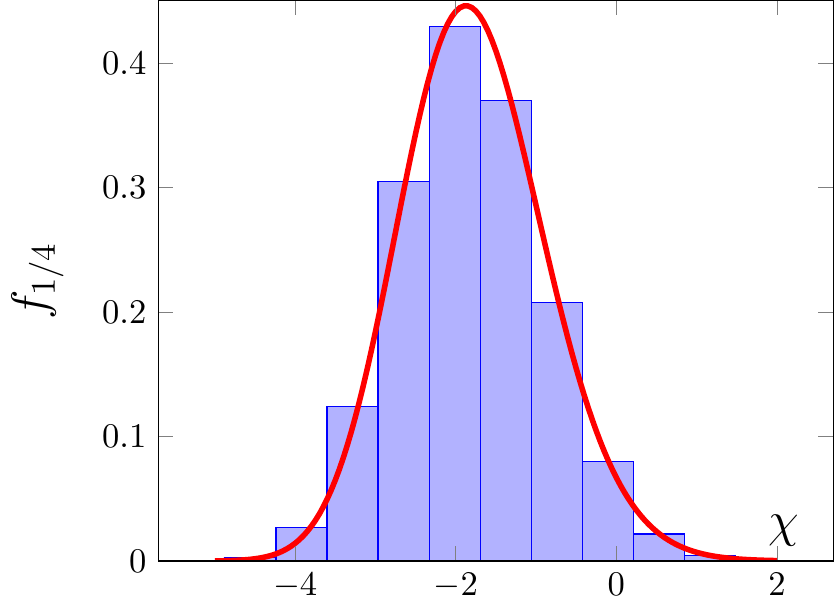}~~
\includegraphics[width=.475\textwidth]{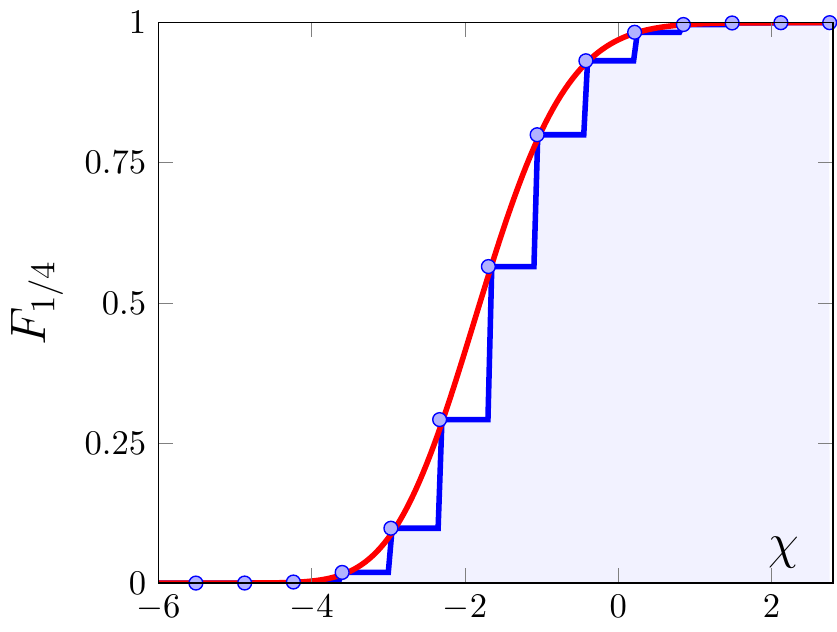}
\caption{\textbf{Left.}  Distribution of the random variable $\mathcal{Z}(\eta)$ defined in Eq.~\eqref{eq:chi_def} calculated from the MC simulation at $\eta=1/4$ and $N=256$ for $\Delta=0$. The red curve is the TW distribution $\mathcal{F}_2'(\chi)$. There are no free parameters in the comparison.~\textbf{Right.} The empirical cumulative distribution of the random variable $\mathcal{Z}_0(\eta)$ is plotted together with the cumulative distribution  $\mathcal{F}_2$ at $\eta=1/4$ and $N=256$ for $\Delta=0$.}
\label{fig:0_bm}
\end{figure}
%%%%%%%%%%%%%%%%%%%%%%%%%%%%%%%%%%%%%%%%%%%%%%%%%%%%%%%%%%%%%%%%%%%%%%%%%%%%%%%%%%%%%%%%%%%% 
\begin{equation}
\label{eq:fdef}
f_{\eta}(\chi)=\Gamma(\eta)N^{1/3}h_{\eta}\bigl(\lfloor\Gamma(\eta)N^{1/3}\chi+N\Lambda(\eta)\rfloor\bigr).
\end{equation}
Upon substituting  Eqs.~\eqref{eq:lambda} and~\eqref{eq:gamma} into Eq.~\eqref{eq:fdef}, one  compares the MC data at finite $N$ with the TW distribution $\mathcal{F}_2'(\chi)$. The result of this comparison is displayed on the left of Fig.~\ref{fig:0_bm} for $\eta=1/4$ and $N=256$. It shows a convincing agreement already for such a value of the lattice size; it should be remarked that there are no free parameters. 

From a statistical viewpoint, it is possible to verify if the random variable $\mathcal{Z}$  is distributed according to $\mathcal{F}_2'$, by performing a Kolmogorov-Smirnov  test~\cite{NPS}. The latter requires to construct the empirical cumulative distribution of $\mathcal{Z}$, which  will be denoted by $F_{\eta}(\chi)$.
On the right panel of Fig.~\ref{fig:0_bm} is plotted $F_{\eta}(\chi)$ at $\eta=1/4$ and $N=256$ together with the cumulative distribution of the TW distribution, denoted by $\mathcal{F}_2(\chi)$.  The test statistics~\cite{KS1, KS2} is  $D_n:=\sqrt{n}~\text{sup}_{\chi}|F_{\eta}(\chi)-\mathcal{F}_2(\chi)|$, where $n$ is the number of points in Fig.~\ref{fig:0_bm}.
For $n=14$, we obtain $D_{14}=1.021\dots$, which corresponds to a $p$-value of about $0.2$~\cite{Vrbik}, well above the conventionally accepted threshold of $0.1$.

\section{Numerical study of the cumulants for $0\leq \Delta<1$ and $\eta=0$}
\label{sec:res_num}
\begin{figure}
\centering
\includegraphics[width=0.48\textwidth]{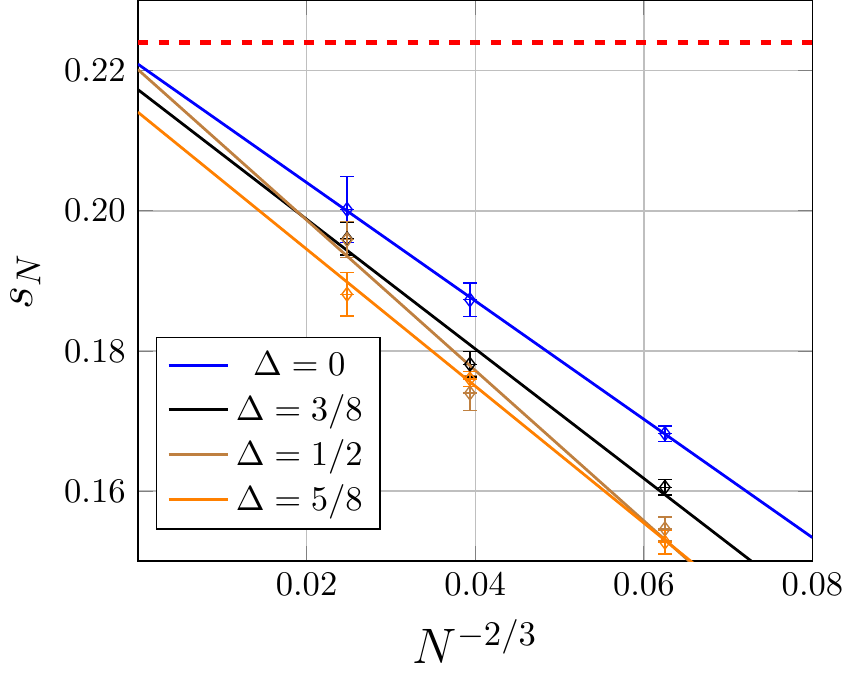}
\includegraphics[width=0.48\textwidth]{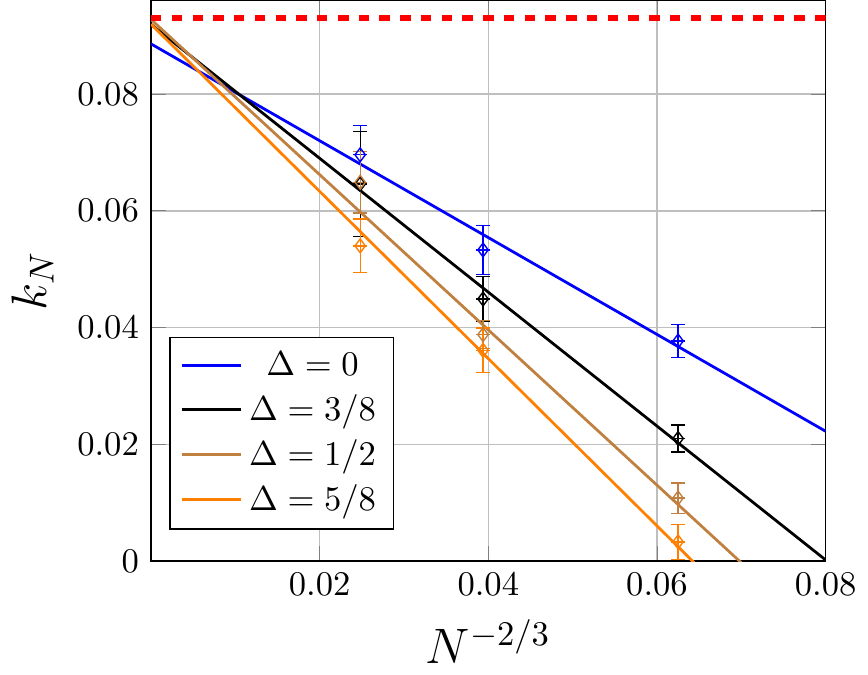}
\caption{\textbf{Left.} Numerical data for the skewness of the empirical distribution $X_{ext}(\eta=0)$ as a function of $N^{-2/3}$ for $N=64,128,256$. The intercept of the regression line is the skewness in the thermodynamic limit. The dashed red line is the TW value at $\beta=2$ which equals $0.224\dots$. \textbf{Right.} Numerical data for the excess kurtosis of the empirical distribution $X_{ext}(\eta=0)$ as a function of $N^{-2/3}$ for $N=64,128,256$. The intercept of the regression line is the excess kurtosis in the thermodynamic limit. The dashed red line is the TW value at $\beta=2$ which equals $0.093\dots$.}
\label{fig:skew}
\end{figure}

In this section we report numerical results for the fluctuations of the random variable $X_{ext}(\eta=0)$, i.e. along the diagonal of the square lattice for values of the anisotropy in the domain $0\leq\Delta<1$. Details about the construction of the empirical probability distribution of fluctuations of the extremal lattice path are given in the Appendix~\ref{app_dist}.
To lighten the notation we will drop the $\eta$-dependence in the remaining part of the discussion and always assume $\eta=0$.
At large $N$, and for any $|\Delta|<1$ the fluctuations of $X_{\text{ext}}$ shall obey (see Eq.~\eqref{eq:chi_def}) 
\begin{equation}
\label{scalingTW}
X_{ext}\stackrel{N\gg 1}{\longrightarrow}N\Lambda+\Gamma N^{1/3}\mathcal{Z},
\end{equation}
where $\mathcal{Z}$ is a TW random variable and the parameters $\Lambda$ and $\Gamma$ are non-universal and $O(1)$. In particular $\Lambda$ can be estimated from the mean of $X_{ext}$ and $\Gamma$ from its variance. 

It should be understood  that Eq.~\eqref{scalingTW} is an asymptotic expansion valid for large enough $N$. In particular, the coefficients $\Lambda$ and $\Gamma$ are the $N\rightarrow\infty$ limit of non-trivial functions of the lattice size, which are not known analytically for $\Delta\not=0$. 
The numerical data will be then extrapolated to $N\rightarrow\infty$ by assuming the same functional form for the finite size-corrections of the distribution cumulants discussed in~\cite{Stephan19 }. A numerical study of $\Lambda$ has been perfomed in detail in Ref.~\cite{LKV2017} and will be not repeated here.
\newline
\begin{table}[t]
\centering
\begin{tabular}{|r||l|l|l|l|}
\hline
 & $\Delta=0$   & $\Delta=3/8 $ & $\Delta=1/2 $ & $\Delta=5/8 $ \\
\hline \hline
$s$ & $0.221(5)$ & $0.217(4)$ & $0.220(4)$ & $0.214(4)$ \\
\hline
$k$ & $0.089(2)$ & $0.092(2)$ & $0.093(3)$ & $0.092(2)$ \\
\hline
\end{tabular}
\caption{The regression estimates for the parameter $s$ and $k$. The number in the bracket is the estimated error in the last digit. The TW values for $s$ and $k$ are $0.224\dots$ and $0.093\dots$ respectively.}
\label{table_sk}
\end{table}
\textit{Skewness and Excess Kurtosis.---}Since the random variable $X_{ext}$ is related to the TW random variable $\mathcal{Z}$ in Eq.~\eqref{scalingTW} by a linear (affine) transformation, cumulants of order larger than two such as the skewness and excess kurtosis, are the same for the two distributions. Their study~\cite{Takeuchi} provides a convenient parameter free test of the universality conjecture and will be considered first. We denote then by $s_{N}$ and $k_{N}$ the empirical skewness and excess kurtosis obtained from the Monte Carlo data for $X_{ext}$ at \textit{finite} $N$. Following Ref.~\cite{Stephan19 } we assume that the leading finite-size correction for large $N$ to the cumulants is $O(N^{-2/3})$, namely
\begin{equation}
\label{scaling_cum}
 s_{N}\stackrel{N\gg 1}{\longrightarrow} s+aN^{-2/3}+o(N^{-2/3});~~k_{N}\stackrel{N\gg 1}{\longrightarrow} k+bN^{-2/3}+o(N^{-2/3}),
\end{equation}
where $o(N^{-2/3})$ represents quantities that vanish faster than $N^{-2/3}$ for $N\rightarrow\infty$. The exponent $2/3$ of the first subleading correction in Eq.~\eqref{scaling_cum} was proposed in 
Ref.~\cite{Stephan19 } in the context of the quantum XXZ spin chain when analyzing fluctuations of the density profile of the interacting fermions close to the edge. Albeit the numerical methods  (Ref.~\cite{Stephan19 } employed DMRG techniques) and the systems are different, the convergence patterns of cumulants toward the thermodynamic limit are resemblant.  

\begin{figure}
 \centering
\includegraphics[width=0.7\textwidth]{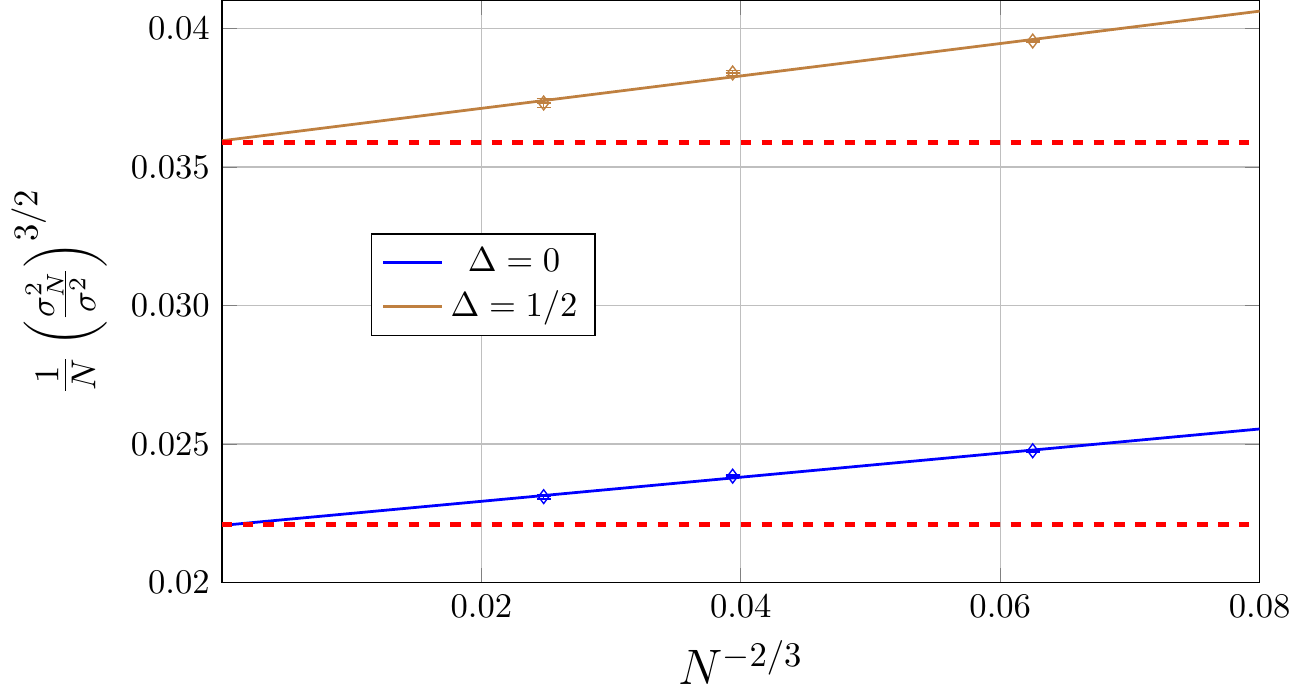}
\caption{Numerical estimation of the coefficient $\Gamma^3$ in Eq.~\eqref{scalingTW} from the variance of the empirical distribution of $X_{ext}$ along the lattice diagonal ($\eta=0$). The predictions at $\Delta=0$ and $\Delta=1/2$ are the $y$-intercepts of the regression lines, see Eq.~\eqref{eq:fs_variance} and perfectly agree with the theoretical values represented by the dashed lines, see main text.}
\label{fig:variance}
\end{figure}

A linear fit of the numerical data for $s_N$ and $k_N$  against $N^{-2/3}$ is shown in Fig.~\ref{fig:skew} by taking into account points at $N=64,128,256$ for $\Delta=0,\frac{3}{8},\frac{1}{2}$ and $\frac{5}{8}$.   The numerical estimates for the coefficients $s$ and $k$ in Eq.~\eqref{eq:chi_def} correspond then to  $y$-intercept of the regression lines; they are collected in Tab.~\ref{table_sk} together with their statistical uncertainties.  The red dashed line in Fig.~\ref{fig:skew} is the TW value. The agreement is convincing although not prefect for the skewness at $\Delta\not=0$ and $\Delta\not=1/2$. The data points in Fig.~\ref{fig:skew} and their error bars are obtained by averaging over several Monte Carlo histograms (5-10 histograms). 
\newline
\textit{Variance at the ice point.---} Let us denote by $\sigma_{N}^2$ the variance of the emprical distribution $X_{ext}$ at \textit{finite} $N$ and let $\sigma^2$ the variance of the TW distribution. From Eq.~\eqref{scalingTW} and the same scaling conjecture proposed by Ref.\cite{Stephan19 } for the higher cumulants, it follows
\begin{equation}
\label{eq:fs_variance}
 \left(\frac{\sigma_N^2}{\sigma^2}\right)^{3/2}\stackrel{N\gg 1}{\longrightarrow}N\Gamma^3+cN^{1/3}+o(1).
\end{equation}
In Sec.~\ref{sec:Delta0}, we recalled the expression for the parameter $\Gamma^3$ along the lattice diagonal at $\Delta=0$, namely $\left.\Gamma^3\right|_{\Delta=0}=\frac{1}{32\sqrt{2}}$ see Eq.~\eqref{eq:gamma} for $\eta=0$. More recently, Ref.~\cite{PS_23} proposed a phenomenological (and not rigorous) formula for $\Gamma^3$ at arbitrary values of $\Delta$, which involves the second derivative of the arctic curve evaluated at the intersection point with the lattice diagonal. At $\Delta=1/2$, the so-called ice point, the arctic curve is an ellipse~\cite{PC} and the parameter $\Gamma^3$ was conjectured to be~\cite{PS_23}: $\left.\Gamma^3\right|_{\Delta=1/2}=\frac{1}{16\sqrt{3}}$. The Ansatz elaborated in~\cite{PS_23} relies on the KPZ~\cite{KPZ} scaling theory applied to the fluctuations of the extremal lattice path.
Fig.~\ref{fig:variance} shows a numerical estimation of the parameter $\Gamma^3$ obtained from a linear regression analogous to that performed for the skweness and excess kurtosis in the previous paragraph. The agreement with the exact~\cite{ADSV2016} and conjectured~\cite{PS_23} expressions at $\Delta=0$ and $\Delta=1/2$ is prefect. An equivalent result can be obtained by averaging the distribution of the extremal lattice path along lines parallel  to the lattice diagonal and requiring best fit of the Monte Carlo data with the TW distribution, see Appendix~\ref{app_average}.

\section{Conclusions}
\label{conclusion}
In this paper, we analyzed numerically, with Monte Carlo techniques, the fluctuations of the extremal lattice path in the six vertex model with DWBC for values of the anisotropy parameter in the domain $0\leq \Delta<1$. Exact results at the free fermion point ($\Delta=0$) show that the intersection of the furthest lattice path with the lattice diagonal follows the Tracy-Widom distribution. The same is true along any line parallel to the diagonal, as long as the contact point~\cite{Gorin} of the arctic curve with the lattice is avoided.
Our numerical simulations were tested thoroughly against these analytic results, showing convincing agreement already for lattices with side-length $N=256$ and without fitting parameters.

Further, we provided strong numerical evidence that the fluctuations of the extremal lattice path follow the Tracy-Widom distribution also for values of the anisotropy parameter in the domain $0\leq \Delta<1$. In particular, we verified the consistency of the scaling hypothesis in Eq.~\eqref{eq:chi_def} by determining the skewness and excess kurtosis of the distribution of $X_{ext}$ along the lattice diagonal. We also verfied, see Eq.~\eqref{eq:fs_variance}, that the width of the fluctuations about the arctic curve is of order $O(N^{1/3})$ with a prefactor that could be extrapolated from the variance of the Monte Carlo data and agrees perfectly with the exact value at $\Delta=0$ and the recent conjecture in~\cite{PS_23} at $\Delta=1/2$.

Our finite size scaling analysis requires relatively small lattices and samples of $O(10^6)$. Therefore it is computationally less demanding than other Monte Carlo attempts to uncover the Tracy-Widom distribution~\cite{MendlS}.   The results are fully consistent with those that Ref.~\cite{Stephan19 }  obtained in quantum spin chains through DMRG simulations.

Our conclusions should extend also to cases $-1<\Delta\leq 0$, which we did not analyze in this paper but belong to the same disordered phase~\cite{BLbook} of the model. Finally, it would be interesting to study fluctuations of the boundary of the inner phase separation curve in the antiferromagnetic regime, see for instance~\cite {CJ2016} or understand the statistics of interior lattice paths~\cite{Ruelle2}. To circumvent the problem of exponentially slow mixing of the Markov chain Monte Carlo dynamics for large and negative $\Delta$, one might implement the algorithm proposed recently in~\cite{Cai}.

\section*{Acknowledgments} 
We thank Alexander Abanov, Filippo Colomo, Giacomo Gori, Andrei Pronko, Herbert Spohn and Jean-Marie St\'ephan for enlightening discussions and interest in this work.  JV is especially grateful to Filippo Colomo and  Jean-Marie St\'ephan for a careful reading of the manuscript and Herbert Spohn for sharing his unpublished draft on the subject.

\section*{Appendices}

\appendix

\section{Numerical evaluation of the distribution of $X_{ext}(\eta)$ for $|\eta|<\frac{1}{2}$}
\label{app_dist}
In this Appendix, we discuss in detail the numerical algorithm that we used to construct the empirical distribution $X_{ext}(\eta)$ for $|\eta|<\frac{1}{2}$.

  For simplicity, we will consider only the isotropic case $a = b$. 
Consider the $1+k_{\max}$ parallel lines $y=y_k$ with $y_k=k/2$ and $k=0,1,\dots,k_{\max}$, see Fig.~\ref{fig:extermal}, and define, see Eq.~\eqref{eq:etadef},
\begin{equation}
\label{discrete_eta}
\eta_k:=2y_k/N=k/N.
\end{equation}

As discussed before, the
parameter $k_{\max}$ must be below $N/2$, in this paper we chose $k_{\max} = N/4$. We begin by describing how the distribution of the random variable $X_{ext}$,  along the line $y=0$, that is the lattice diagonal, will
be constructed. We start with a random lattice state $S^{(0)}$ and an $N/2$-dimensional vector $v_0^{(0)}$:
\be
v_0^{(0)} = (0, ~ 0, ~ 0, ~ \dots, ~ 0).
\label{h0}
\ee   
The state $S^{(0)}$ will be used to obtain a new vector $v_0^{(1)}$ in the following way. We move from each of the four corners towards the center of the lattice
until the first vertex with a state different from the state at the corner is met. The integer variable $x$ will be used to count the steps. For example, beginning at the NW corner at $x=0$, we walk along the diagonal
until, at step $x = x_{{\rm NW}}$, we reach a vertex which has a state different from $a_1$. Beginning instead at the SW corner,
we continue until we reach a vertex which does not have state $b_2$; and so on. The vector $v_0^{(1)}$ will then have the entries
\be 
v_0^{(1)}(x) = v_0^{(0)}(x) + \sum_{\iota \in \{ {\rm NE}, ~{\rm NW}, ~{\rm SW}, ~{\rm SE}\}} \delta_{x, ~ x_{\iota}}; ~ x = 0, ~ 1, ~ \dots, ~ N/2 - 1. 
\label{h02}
\ee
Next, we perform one MC sweep, see Sec.~\ref{sec:num}. This will generate a new, independent random lattice
state $S^{(1)}$. The process is then repeated and we will obtain a new vector $v_0^{(2)}$ from the vector $v_0^{(1)}$ and the state $S^{(1)}$. Altogether $T-1$ sweeps will be made. The final vector $v_0^{(T-1)}$ will thus be obtained from $T$ states $S^{(0)}, ~S^{(1)}, ~ \dots , ~S^{(T-1)}$. It is convenient to use the opposite direction, from the center of the lattice towards the corner. To this end, we define the normalized histogram $h_0$, see Sec.~\ref{sec:num}, as follows
\be
h_0(x) := v_0^{(T-1)}\bigl(N/2 - 1 - x\bigr)/4T; ~ x = 0, ~ 1, ~ \dots, ~ N/2 - 1. 
\label{h03}
\ee
The distribution obtained from Eq.~\eqref{h03} is illustrated on the upper left side of Fig.~\ref{fig:hist} for $\Delta=1/2$ and $N=256$. Here $x$ is the distance from the center along the diagonal, which in the coordinate systems of Fig.~\ref{fig:extermal}  coincides indeed with the value of the $x$-coordinate in the NW corner. As mentioned in Sec.~\ref{sec:intro}, the diagonal distance between two neighboring vertices is $1$. In this paper, $T$ will have the value $T=10^6$ throughout.

In order to provide evidence of the validity of the scaling hypothesis in Eq.~\eqref{scalingTW} also at $\eta\not=0$ and away from the free fermionic point, we construct  $k_{\max}$ histrograms of $X_{ext}(\eta\not=0)$, i.e. along lines parallel to the lattice diagonal. The vectors $\{v_{\eta_k}^{(T-1)}\}_{k=1}^{k_{\max}}$ are built taking into account that the process is begun on side of the lattice and not in the corner. Therefore,
there are eight simultaneous processes and not four. We thus obtain another set of 
vectors $\{v_{\eta_k}^{(T-1)}\}_{k = 1}^{k_{\max}} $. From these, we will obtain normalized, reversed histograms
$\{ h_{\eta_k} \}$ if, for each $k=1,\dots, k_{\max}$, we define $h_{\eta_k}$ as
\be
h_{\eta_k}(x) := v_{\eta_k}^{(T-1)}(N/2 - 1 - x+y_k)/8T; \quad x = -y_k,\dots, ~ N/2 - 1-y_k.
\label{hk}
\ee
The  histogram $h_{1/4}(x)$ obtained when $k=k_{\max}=N/4$ is displayed for $\Delta=1/2$ and $N=256$ on the bottom left of Fig.~\ref{fig:hist}.  
\section{Distribution average over lines parallel to the lattice diagonal}
\label{app_average}
\begin{table}[t]
\centering
\begin{tabular}{|r||l|l|l|l|}
\hline
$N $  & $\Delta=0$   & $\Delta=3/8 $ & $\Delta=1/2 $ & $\Delta=5/8 $ \\
\hline \hline
$64$ & $1.822(13)$ & $2.465(12)$ & $2.849(14)$ & $3.47(3)$ \\
\hline
$128$ & $3.348(12)$ & $4.557(22)$ & $5.32(4)$ & $6.47(3)$ \\
\hline
$256$ & $6.29(7)$  & $8.60(14)$ & $10.05(16)$  & $12.46(24)$ \\
\hline
\end{tabular}
\caption{The parameter $[\Gamma_N]^3$ as a function of $N$ and $\Delta$. The values of $N$ are contained in the leftmost column and the values of $\Delta$ are shown in the uppermost row. The number in the bracket is the estimated error in the last digit.}
\label{table1}
\end{table}
We assume that all the histograms $h_{\eta_k}$ obtained in Appendix~\ref{app_dist} once shifted and rescaled to have the same mean and variance converge for large $N$ to the same distribution. To test the scaling hypothesis in Eq.~\eqref{eq:chi_def}, we then average over them as described in the following.
 
First one  calculates the mean $\mu_k$ and the variance $\sigma_k$ of the distributions $h_{\eta_k}$ for all $k=0,\dots,k_{\max}$. We have
\begin{equation} 
\label{eq:mean_variance}
\mu_k = \sum_{x = -y_k}^{N/2 - 1-y_k} x~h_{\eta_k}(x),\quad
\sigma_k^2 = \sum_{x = -y_k}^{N/2 - 1-y_k}(x - \mu_k )^2 ~ h_{\eta_k}(x).
\end{equation}
At this point, we will leave the actual lattice and define our distributions in a new, one-dimensional space. To perform with greater numerical accuracy the averaging procedure, it is convenient to magnify the bin width by a constant factor $\theta$, such that {$1<\theta< k_{\max}$}. In terms of the new variable $t=\theta x$, the probability distributions in Eq.~\eqref{hk} are then~\footnote{{The entries of histograms are now indexed by $\theta k_{\max}+1$ integers and, provided they have the same mean and variance,  can be summed without the need of defining a piecewise function on the reals. The optimal value of $\theta$ is chosen in order to to render the averaged histogram as smooth as possible. Its effect on the fluctuations of the random variable $X_{ext}$ can be trivially traced back, see Eq.~\eqref{eq:TW_conj}.}}
\begin{equation}
\label{eq:hprimedef}
h^{\sharp}_{\eta_k}(t)=\frac{1}{\theta}h_{\eta_k}\left(\left\lfloor\frac{t}{\theta}\right\rfloor\right).
\end{equation}
The value $\theta = 8$ will be used in this paper. By following the assumption stated at the beginning of this section, we further perform for all $k=0,\dots,k_{\max}$ the linear change of variable
\begin{equation}
\xi=\frac{\sigma_0}{\sigma_k}(t-\theta\mu_k),
\end{equation}
which transforms the probability distributions of Eq.~\eqref{eq:hprimedef} into
\begin{equation}
\label{eq:pdf_final}
h_{\eta_k}^{\flat}(\xi)=\frac{\sigma_k}{\theta\sigma_0}h_{\eta_k}\left(\left\lfloor\frac{\sigma_k}{\sigma_0\theta}\xi+\mu_k\right\rfloor\right).
\end{equation}
It is easy to show, by recalling Eq.~\eqref{eq:mean_variance}, that all the probability distributions in Eq.~\eqref{eq:pdf_final} have zero mean and variance $(\theta\sigma_0)^2$. According to our working hypothesis, they are then different histogram representations of the same probability distribution of the random variable
\begin{equation}
\label{eq:xidef}
\Xi:=\theta X_{ext}(\eta=0)+C.
\end{equation}
In Eq.~\eqref{eq:xidef}, $C$ is a constant that ensures that $\Xi$ has zero mean. We estimate the probability distribution of the random variable $\Xi$ with the average
\begin{equation}
\label{eq:average}
\bar{h}_{0}(\xi):=\frac{1}{1+k_{\max}}\sum_{k=0}^{k_{\max}}h^{\flat}_{\eta_k}(\xi);
\end{equation}
in Eq.~\eqref{eq:average} the variable $\xi$ belongs to the union of all the support of the histograms $h_{\eta_k}$ in Eq.~\eqref{eq:pdf_final}. 

\begin{figure}[t]
\centering
\includegraphics[width=0.48\textwidth]{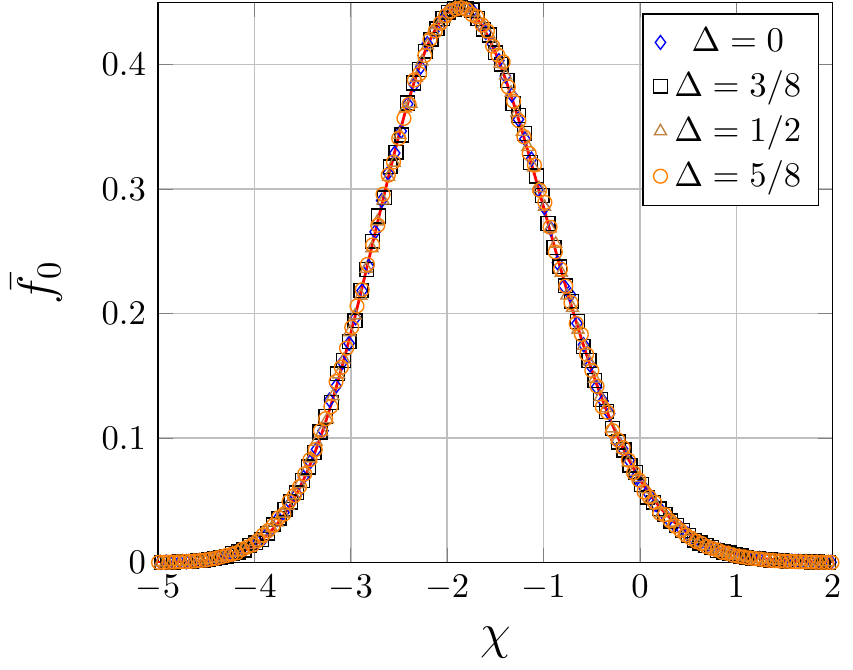}
\includegraphics[width=0.48\textwidth]{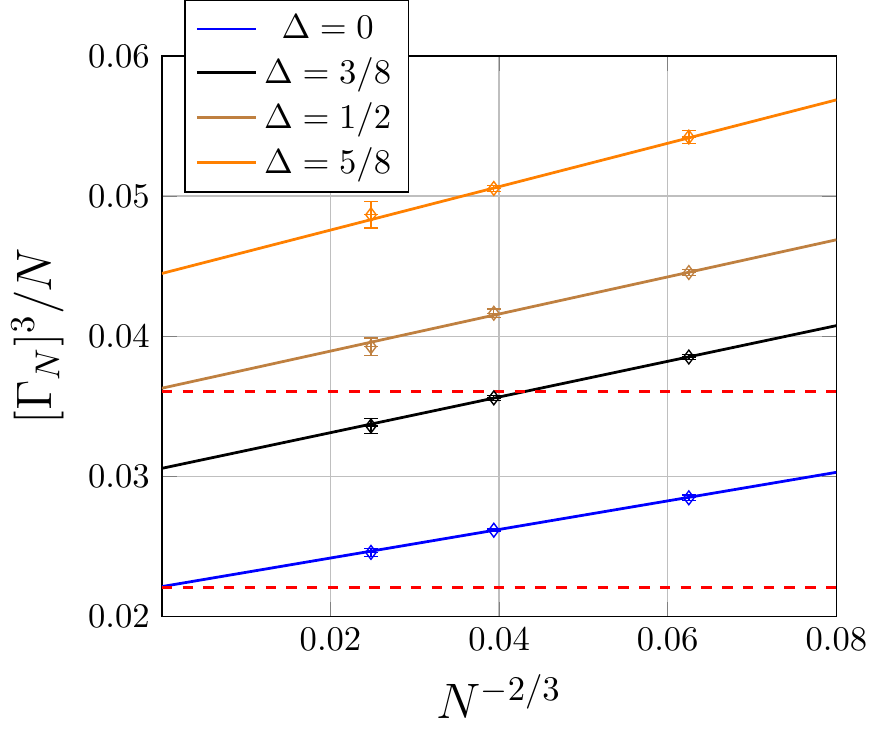}
\caption{\textbf{Left.} Numerical data for the distribution $\bar{f}_0(\chi)$ in Eq.~\eqref{eq:f0def} at different values of $\Delta$ and $N=256$. The parameters $\Gamma_N$ and $\Lambda_{\beta}$ are obtained by requiring a best fit of the data with the TW distribution, which is the red curve underneath. \textbf{Right.~} Numerical data for $[\Gamma_N]^3/N$ from Tab.~\ref{table1} together with the regression lines, see Eq.~\eqref{eq:finitesizealpha}.  The red dashed lines are the analytical results for the $y$-intercept at $\Delta=0$ and $\Delta=1/2$,  Sec.~\ref{sec:res_num}. }
\label{fig:TW_all}
\end{figure}

Instead of calculating  the cumulants as in Sec.~\ref{sec:res_num}, one could  test the scaling hypothesis in Eq.~\eqref{eq:chi_def} (and Eq.~\eqref{scalingTW}) by directly fitting  Eq.~\eqref{eq:average} with the TW distribution.
In particular, recalling Eq.~\eqref{eq:xidef}, we analyze weather 
\begin{equation}
\label{eq:TW_conj}
\bar{h}_0(\xi)\stackrel{N\gg 1}{\longrightarrow}\frac{1}{\Gamma_N\theta}\mathcal{F}_2'\left(\frac{\xi-C-\theta\Lambda_N}{\Gamma_N\theta}\right),
\end{equation}
with $\mathcal{F}_2'$, the TW distribution, by fitting at \textit{finite} $N$ the two parameters $\Lambda_N$ and $\Gamma_N$ above. As also discussed already in Sec.~\ref{sec:res_num}, $\Lambda_N$ and $\Gamma_N$ are non-trivial and unknown functions of the lattice size $N$ such that (see Eq.~\eqref{eq:chi_def})
\begin{equation}
 \lim_{N\rightarrow\infty}\frac{\Lambda_N}{N}=\Lambda(\eta=0),\quad\lim_{N\rightarrow\infty}\frac{\Gamma_N}{N^{1/3}}=\Gamma(\eta=0).
\end{equation}
Working with $N$ up to $256$, it becomes crucial to evaluate the leading finite size corrections to those parameters, in particular following~\cite{Stephan19 } one has, see Sec.~\ref{sec:res_num},
\begin{equation}
\label{eq:finitesizealpha}
 \frac{[\Gamma_N(0)]^3}{N}\stackrel{N\gg 1}{\longrightarrow}[\Gamma(0)]^3+cN^{-2/3}+o(N^{-2/3}). 
\end{equation}
In Tab.~\ref{table1} are collected the numerical estimates for $[\Gamma_N(0)]^3$ obtained by requiring best fit of the Monte Carlo data for the averaged distribution of the extremal lattice path with the TW distribution (see Eq.~\eqref{eq:TW_conj}).
A linear fit of the numerical data for $[\Gamma_N]^3/N$ given in Tab.~\ref{table1} against $N^{-2/3}$ is shown on the right panel of Fig.~\ref{fig:TW_all} by taking into account points at $N=64,128,256$.   The numerical prediction for the coefficient $[\Gamma(0)]^3$ in Eq.~\eqref{eq:chi_def}  corresponds then to  $y$-intercept of the regression lines.  As in Sec.~\ref{sec:res_num}, see Fig.~\ref{fig:variance}, the red dashed lines are the exact~\cite{ADSV2016} and conjectured values~\cite{PS_23} at $\Delta=0$ and $\Delta=1/2$ respectively. The agreement is perfect also with this approach.

Finally, Eq.~\eqref{eq:TW_conj} is equivalent to the statement that for large enough $N$ 
\begin{equation}
\bar{f}_{0}(\chi):=\Gamma_N\theta~\bar{h}_{0}\bigl(\Gamma_N\theta~\chi+\theta\Lambda_N+C\bigr)
\label{eq:f0def}
\end{equation}
converges to the TW distribution for  any $0\leq \Delta<1$. On the left of Fig.~\ref{fig:TW_all}, we show on the same plot the numerical results for  $\bar{f}_{0}(\chi)$ at $N=256$ and $\Delta=0,~\frac{3}{8},~\frac{1}{2},~\frac{5}{8}$. The continuous red curve underneath, which is barely visible, is the TW distribution $\mathcal{F}_2'(\chi)$.

%%%%%%%%%%%%%%%%%%%%%%%%%%%%%%%%%%%%%%%%%%%%%%%%%%%%%%%%%%%%
%%%% Background 
%%%%%%%%%%%%%%%%%%%%%%%%%%%%%%%%%%%%%%%%%%%%%%%%%%%%%%%%%%%%

\end{document}